\PassOptionsToPackage{usenames,dvipsnames,svgnames,table}{xcolor}
\documentclass[sigconf,screen,nonacm]{acmart}

\setcopyright{none}
\renewcommand\footnotetextcopyrightpermission[1]{}

\usepackage{xspace}

\usepackage{booktabs}
\usepackage{enumitem}

\graphicspath{{figures/}{./figs/}}

\setlist{nosep,leftmargin=*}
\usepackage{microtype}
\usepackage{xspace}
\newcommand{\system}{\textsc{HandMade}\xspace}

\definecolor{quote-gray}{gray}{0.3}

\title{\system: Spatial Prompting for Generative 3D Creation with Part-Labeled VR Sketches}

\author{Jialin Huang}
\authornote{Corresponding author.}
\orcid{0009-0000-9915-1252}
\email{jhuang26@gmu.edu}
\affiliation{%
  \institution{George Mason University}
  \city{Fairfax}
  \country{United States of America}
}

\author{Rana Hanocka}
\orcid{0000-0003-3214-3703}
\email{ranahanocka@uchicago.edu}
\affiliation{%
  \institution{University of Chicago}
  \city{Chicago}
  \country{United States of America}
}

\author{Ariel Shamir}
\orcid{0000-0001-7082-7845}
\email{arik@runi.ac.il}
\affiliation{%
  \institution{Reichman University}
  \city{Herzliya}
  \country{Israel}
}

\author{Yotam Gingold}
\orcid{0000-0002-5381-2104}
\email{ygingold@gmu.edu}
\affiliation{%
  \institution{George Mason University}
  \city{Fairfax}
  \country{United States of America}
}

\begin{document}

\begin{teaserfigure}%
\centering
\includegraphics[width=\textwidth]{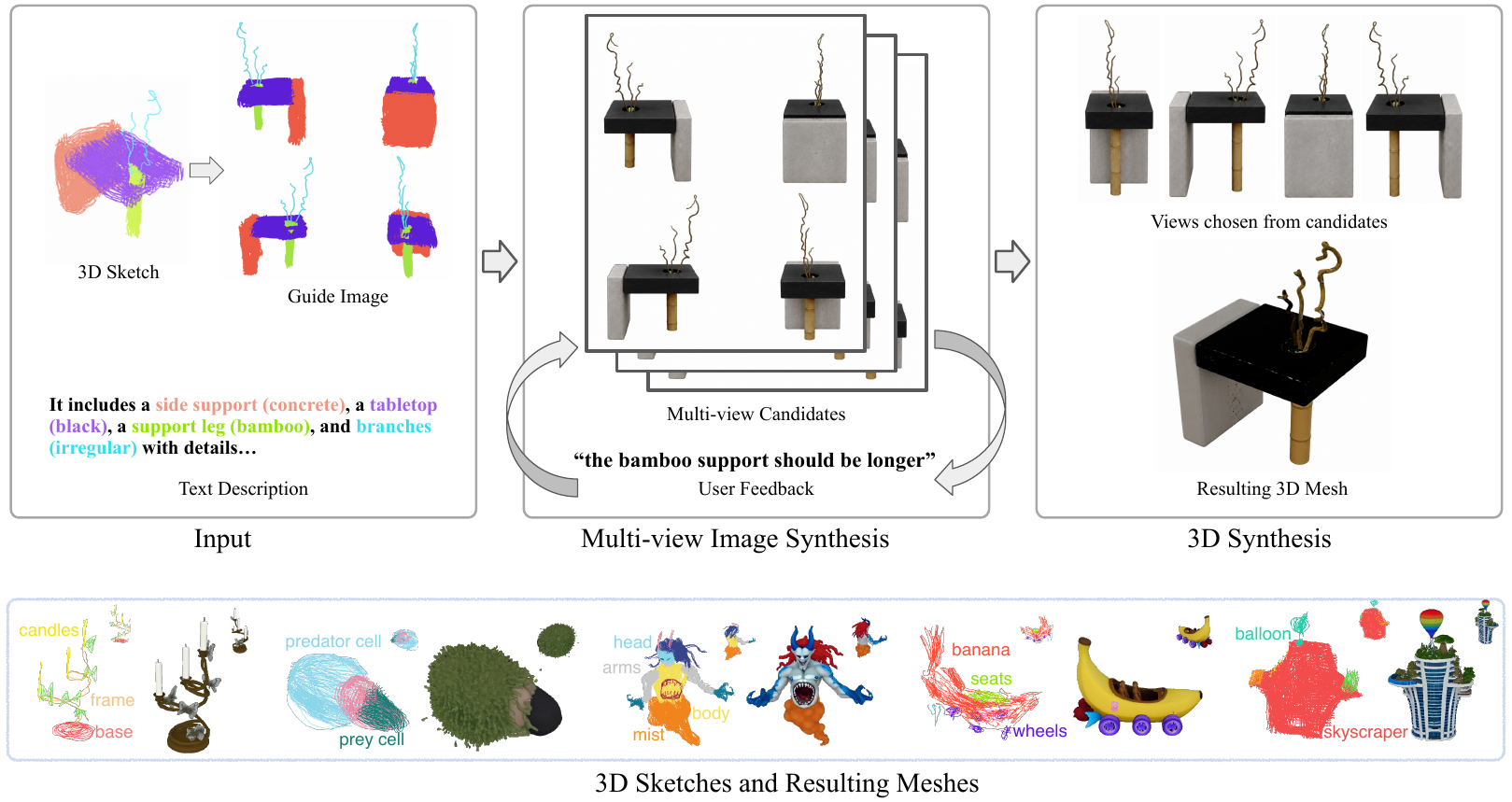}
\caption{Above: Overview of the \system pipeline.
A rough, part-labeled VR 3D sketch is rendered into a multi-view guide image.
The guide and object- and part-level text descriptions condition multi-view image synthesis.
Users can revise candidates through conversational feedback and select views from one or more candidates.
The selected views are then used to reconstruct a textured 3D mesh.
Below: Example annotated 3D sketches and output textured meshes created with \system.
Each sketch part is color-coded and paired with text that names and describes it; only selected labels are shown for readability. Alternate views are shown inset.}
\Description{Above: Three-stage pipeline diagram. Above-Left: a color-coded 3D sketch is rendered into four guide views and paired with a text description of the object and its parts. Above-Center: a stack of multi-view image candidates is connected to a user-feedback loop, illustrated by the request ``the bamboo support should be longer.'' Above-Right: four views selected from the candidates are used to reconstruct a textured 3D mesh with a black tabletop, a concrete side support, a bamboo leg, and branches.
Below: Teaser gallery with five examples. Each example consists of a sketch on the left, a texture mesh on the right, and second-view insets. The sketches depict user-authored 3D sketches segmented into color-coded parts and paired with text that names and describes the colored parts.}
\label{fig:pipeline}
\label{fig:teaser}
\end{teaserfigure}

\begin{abstract}

Text-to-3D generation lowers the barrier to 3D content creation, but text alone offers limited control over spatial relationships and structures. In contrast, using your hands to model 3D objects has been shown to be simple and intuitive, but lacks the ability to be precise and detailed.
We present \system, an authoring system combining rough, part-labeled VR sketches with text for open-domain generation.
Users create and describe a 3D sketch. \system converts it into multiview image guidance for 3D textured mesh synthesis.
Empirical evaluations found closer sketch alignment on average than text-only and 2D sketch baselines.
Users rated VR drawing higher than 2D for expressing structure and part relationships and preferred it for similar future tasks.
Our findings highlight rough 3D sketches as usable generative guidance and intermediate views as checkpoints for selective revision and reuse.

\end{abstract}

\maketitle

\section{Introduction}

Designers as well as novices can have rich and creative ideas for 3D objects, but turning those ideas into editable assets still requires substantial modeling expertise.
Recent generative models have made 3D generation more accessible by using pretrained image diffusion models and multi-view priors as generative backbones \citep{poole2022dreamfusion,lin2023magic3d,wang2023prolificdreamer,chen2023fantasia3d,shi2023mvdream}.
However, text remains an indirect interface for spatial design intent.
A prompt can name an object and describe style, but it is difficult to specify where parts should go, how large they should be relative to each other, how they connect, or what makes a particular design different from a generic instance. In addition, editing or modifying a 3D model using text can also become cumbersome.
These limitations are familiar in HCI work on generative AI: prompt-based workflows can reduce the cost of producing candidates, while also shifting effort to prompt formulation, iteration, evaluation, and repair \citep{wang2024promptcharm,tankelevitch2024metacognitive,weisz2024designprinciples}.
Recent spatial-control work in 3D generation makes a similar argument from the model side: text and images are often insufficiently specific for geometric control, motivating explicit spatial inputs during generation \citep{fedele2026spacecontrol}.

Image-to-3D and multi-view-to-3D systems provide a complementary path.
They can reconstruct plausible 3D assets from images or multi-view conditions \citep{liu2023zero123,liu2023syncdreamer,long2024wonder3d,hong2023lrm,tochilkin2024triposr,xu2024instantmesh,boss2024sf3d,huang2025spar3d,hunyuan3d2025}.
Yet these systems usually assume that the user already has a relatively complete visual reference.
For early-stage asset creation, especially for non-experts, the desired input is often less polished: a rough arrangement of parts, a few words about identity and style, and a willingness to iterate once concrete results appear.

In contrast, sketching is a natural alternative.
Sketch-based modeling has long shown that drawing can support rapid shape ideation, from 2D sketch-to-surface systems such as Teddy \citep{igarashi1999teddy} to 3D curve and freeform-surface interfaces \citep{bae2008ilovesketch,nealen2007fibermesh}.
For 3D generation, VR has the advantage that users can sketch directly in the same space where the object will eventually exist, giving them a way to express rough volume, orientation, part placement, and cross-view relationships \citep{jackson2016liftoff,arora2017sketching,machuca2018multiplanes}.
However, the challenge is that freehand 3D sketches are far from complete 3D models;
they are sparse, incomplete, imprecise, noisy, and often deliberately abstract.
A few strokes can indicate a car body, a volcano, or a playground structure to a human collaborator, but those strokes do not comply with the dense visual evidence expected by current image-to-3D systems and are far outside the category assumptions of many sketch-to-3D methods.

\begin{figure*}[t]
  \centering
  \includegraphics[width=\textwidth]{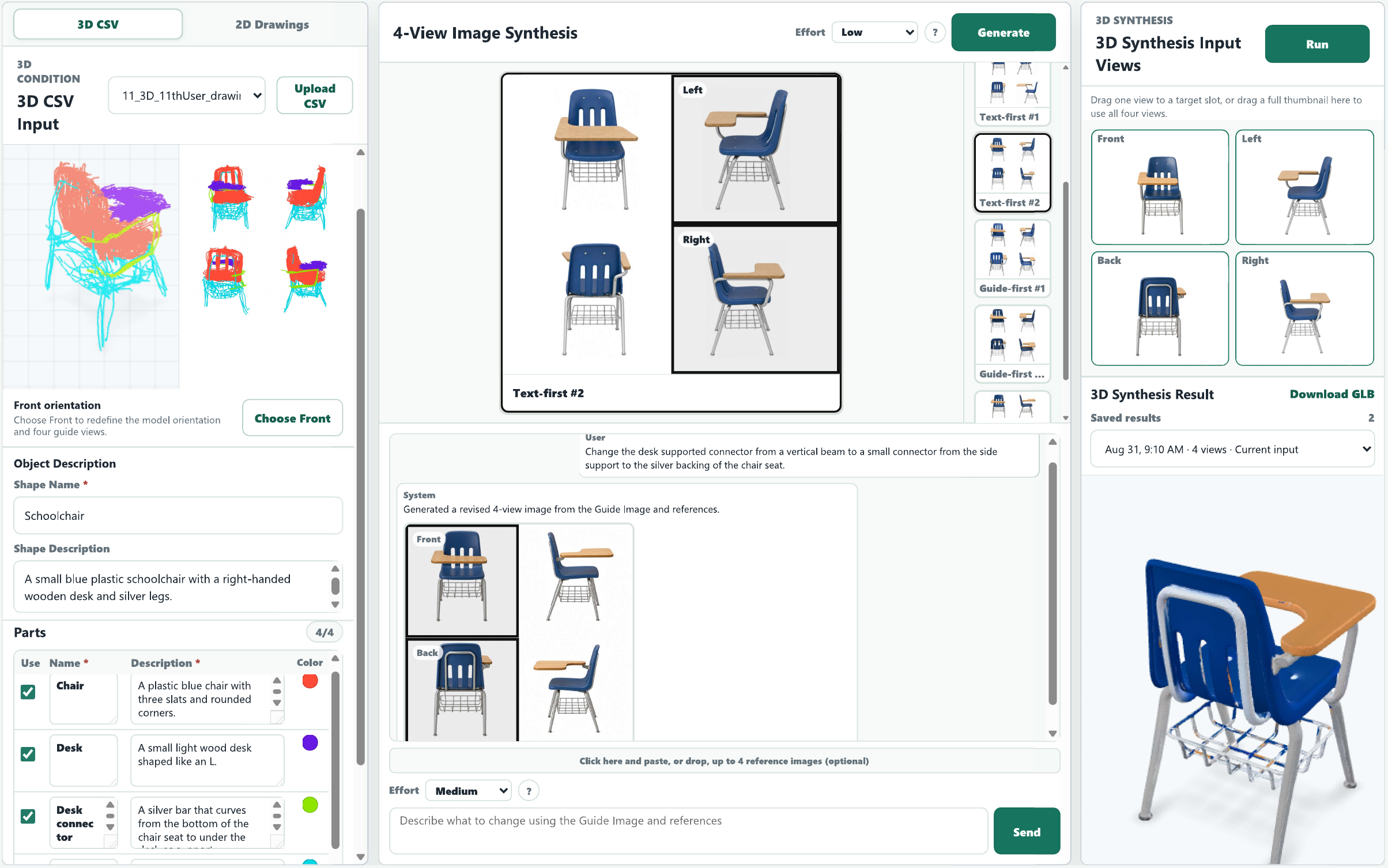}
  \caption{The \system GUI created in response to the pilot study. Left: the 3D sketch preview, rendered guide views, and editable object and part descriptions. Center: multi-view candidate generation and inspection, with a conversational panel below showing a revision request and the resulting multi-view image. Right: view selection for 3D synthesis, followed by mesh inspection and export.}
  \Description{Three-column screenshot of the browser-based GUI. The left column shows a color-coded 3D sketch of a chair, four rendered guide views, and editable object and part descriptions. The center shows a selected four-view chair candidate alongside candidate thumbnails. Below, the conversational panel shows a request to change the writing-tablet connector and the returned four-view image, followed by an area for attaching reference images and a text-entry field. The right column shows front, left, back, and right image slots, a 3D synthesis control, and a textured mesh preview of a blue chair with a wooden writing tablet and a gray metal frame, with an export control.}
  \label{fig:gui}
\end{figure*}

We present \system, a system that combines rough, part-labeled VR 3D sketches with text descriptions for open-domain 3D asset generation and editing.
Users author a rough 3D sketch part-by-part and provide object- and part-level descriptions (Figure \ref{fig:teaser}).
The 3D sketch anchors part placement and proportions and provides multi-view consistency,
while text descriptions supply object and part identities and materials.
\system renders the 3D strokes into part-aware multi-view guidance, while binding sketch colors to part descriptions in a structured prompts. It then generates a detailed multi-view image candidates, and reconstructs selected views as a textured mesh.
Furthermore, \system's interface exposes these intermediate stages for inspection and revision (Figure~\ref{fig:gui}).
\system uses existing image-generation and image-to-3D backbones as modular components and leverages their open-domain generality.

Our central contribution is showing how rough 3D input can serve as a spatial prompt: the pipeline converts sparse 3D intent into visual conditions that general-purpose generative models can use.
Figure~\ref{fig:teaser} shows representative examples, pairing user-authored part-labeled 3D sketches with the textured meshes produced by the pipeline.
We designed and evaluated \system with a 10-participant pilot study, an 18-participant controlled experiment, and a technical evaluation.
We refined \system's authoring interface with feedback from the pilot study.
In the controlled experiment, participants reproduced the same target object using VR 3D drawing and multi-view 2D drawing with an otherwise identical generation workflow.
Empirically, the 3D condition reproduced the input with lower error. Most users preferred 3D drawing over multi-view 2D drawing for creating 3D consistent guidance by construction, despite the greater physical demands.
(The multi-view 2D drawing scenario is still supported by \system as an alternative when VR drawing is unavailable or cannot be used.)
The technical evaluation over 20 varied examples tests how well \system preserves user-authored 3D spatial structure compared with text-only generation and 2D sketch-based baselines.

\section{Related Work}

Generative 3D systems differ not only in model architecture, but also in the kinds of input they expect from users: text prompts, reference images or views, sketches, spatial controls, or combinations of these signals.
This distinction matters for \system because our focus is the translation from rough, part-labeled VR 3D sketches and text descriptions into the cleaner visual conditions that current open-domain 3D generators can use.
Our work is therefore related to 3D generative backbones, multimodal authoring interfaces, sketch-conditioned generation, and VR spatial drawing as an imperfect but expressive input medium.

\subsection{Open-Domain 3D Generation Backbones and Control}

Open-domain 3D generation provides the synthesis backbones that make \system possible, but not the interaction layer needed to express rough spatial intent.
Text-to-3D methods such as DreamFusion~\citep{poole2022dreamfusion}, Magic3D~\citep{lin2023magic3d}, ProlificDreamer~\citep{wang2023prolificdreamer}, Fantasia3D~\citep{chen2023fantasia3d}, MVDream~\citep{shi2023mvdream}, and Instant3D~\citep{li2024instant3d} made open-vocabulary 3D synthesis increasingly practical.
Image- and multi-view-conditioned systems such as Zero-1-to-3~\citep{liu2023zero123}, SyncDreamer~\citep{liu2023syncdreamer}, Wonder3D~\citep{long2024wonder3d}, LRM~\citep{hong2023lrm}, TripoSR~\citep{tochilkin2024triposr}, CRM~\citep{wang2024crm}, InstantMesh~\citep{xu2024instantmesh}, Unique3D~\citep{wu2024unique3d}, SF3D~\citep{boss2024sf3d}, SPAR3D~\citep{huang2025spar3d}, and Hunyuan3D~\citep{hunyuan3d2025} further support fast 3D asset generation from visual conditions.
These systems generally expect text, images, or generated views, whereas early-stage creators may only have a sparse arrangement of parts and a verbal description.
\system therefore treats rough VR sketching as an authoring interface that must be translated into the visual conditions current 3D generators can consume.

Control-oriented 3D generation shows the value of explicit structure, particularly because early creative intent is often compositional: users think in terms of parts, proportions, and spatial relations rather than a single global description \cite{tversky1984objects}.
Several methods introduce structured geometric controls for generation. SpaceControl~\citep{fedele2026spacecontrol} adds ``test-time''
geometric control through primitives or meshes.
ART-DECO~\citep{chen2025artdeco} uses text prompts to add fine details to editable shapes, 
Points-to-3D~\citep{xia2026points3d} uses point cloud priors,
and CompoSE~\citep{slim2026compose} synthesizes and edits part-separated 3D shapes from coarse part boxes.
PartGen~\citep{chen2025partgen} generates a part-segmented 3D shape from a text prompt, image, or existing 3D shape via view-consistent image segmentation.
Some systems operate through editing or procedural representations.
Instant3dit~\citep{barda2025instant3dit}, which we include as a multi-view inpainting/editing baseline in Section~\ref{sec:technical-evaluation}, edits existing 3D assets through multi-view inpainting, while LL3M~\citep{lu2025ll3m} generates editable Blender code.
These systems demonstrate useful forms of structural control,
but their controls are either too precise (accurate and consistent 2D or 3D guidance) or too vague (bounding volumes).
\system instead studies an earlier authoring step, where the user's part-labeled 3D strokes may have incomplete boundaries, inaccurate proportions, or imprecise local shape cues,
and translates rough part-labeled 3D strokes and text into visual conditions for generation.

\subsection{Prompting and Multimodal Generative Interfaces}

Prompting research motivates \system as a multimodal authoring workflow, not only as a model-control technique.
HCI work on generative AI shows that prompts are expressive but cognitively demanding: users must decide what to specify, predict model behavior, inspect outputs, and iterate through revisions~\citep{tankelevitch2024metacognitive,weisz2024designprinciples}.
Earlier work on multimodal interaction similarly shows that speech and pen input are often coordinated rather than redundant, with users distributing intent across modalities in temporally structured ways~\citep{oviatt1999tenmyths,oviatt_integration_1997}.
PromptCharm~\citep{wang2024promptcharm}, for example, supports text-to-image prompting through multimodal refinement and visualization.
DrawTalking~\citep{rosenberg2024drawtalking} and Editing Reality~\citep{weng2026editingreality} are closer to creative spatial authoring; they combine sketching with speech or language so users can build, modify, or control generated content in interactive worlds or mixed-reality environments.
SketchGPT~\citep{huang2025sketchgpt} makes a broader interface argument, using sketch and speech over an application interface so an LLM can infer context-aware user intentions and execute toolkit-aware operations.
JustShape~\citep{duan2026justshape} similarly uses co-speech gestures to complement language for LLM-powered 3D parametric modeling, showing how spatial gestures can make implicit 3D design intent more explicit.
\system follows this broader view of prompting, but uses sketching
for part-labeled 3D spatial input.
DenseDiffusion~\citep{kim2023densediffusion} pairs regional text descriptions with spatial masks and modulates attention to guide where the described content appears in a generated image, providing a precedent for explicitly linking descriptions to image regions.
Unlike these systems, \system binds user-defined part labels to sketched regions, uses text for object identity, material, style, and local details, and translates the combined input into multi-view guidance for open-domain 3D asset generation.

\subsection{Sketch-Conditioned 3D Generation}

Sketch-to-3D methods validate drawing as a compact shape condition, but most assume cleaner and more image-like inputs than rough VR strokes.
Sketch2Model~\citep{zhang2021sketch2model} and Sketch2Mesh~\citep{guillard2021sketch2mesh} reconstruct shapes from single freehand sketches, while SDFusion~\citep{cheng2023sdfusion} and LAS-Diffusion~\citep{zheng2023lasdiffusion} support multimodal or locally controllable generation with learned 3D representations. We include LAS-Diffusion as a 2D sketch-only shape-generation baseline in Section~\ref{sec:technical-evaluation}.
These methods establish sketching as a useful control signal, but their input assumptions differ from \system's setting. Sketch2Model and Sketch2Mesh interpret a single 2D freehand sketch as the primary shape observation, so the sketch is a view-dependent drawing rather than a 3D spatial trace. SDFusion and LAS-Diffusion operate through learned 3D shape representations and dataset-supported priors, which makes them less suited to rough open-domain objects authored from sparse VR strokes. \system instead starts from part-labeled 3D strokes whose part boundaries, proportions, and local shape cues may be incomplete or inaccurate.

Recent work broadens sketch-conditioned generation toward more open or multimodal settings.
Sketch-A-Shape~\citep{sanghi2023sketchashape} studies zero-shot sketch-to-3D generation, Doodle Your 3D~\citep{bandyopadhyay2024doodle} handles abstract part-aware sketches, Control3D~\citep{chen2023control3d} and Sketch2NeRF~\citep{chen2024sketch2nerf} incorporate sketch guidance into text-to-3D optimization, and SKED~\citep{mikaeili2023sked} uses sketches for local text-guided editing.
PASTA~\citep{lee2025pasta} further combines sketches and text through a part-aware, text-aligned representation, using semantic part information to compensate for ambiguous visual cues in a sketch.
SketchDream~\citep{liu2024sketchdream} and MeshPad~\citep{li2025meshpad} are especially relevant because they combine sketch guidance with text-conditioned generation or interactive mesh editing; we include both as sketch-based baselines in Section~\ref{sec:technical-evaluation}.
However, these systems still typically assume a clean 2D sketch, depth or image-like guidance, an existing object, or a dedicated sketch-to-shape model.
\system addresses a different input condition: user-defined part labels on casual strokes drawn directly in 3D, paired with text and converted into multi-view guidance for existing open-domain generators.

Other recent systems reduce ambiguity by strengthening 2D or multi-view drawing inputs.
Magic3DSketch~\citep{zang2026magic3dsketch}, MSU-3D~\citep{zhang2026msu3d}, OmniSketch~\citep{sun2026omnisketch}, From Sketch to Reality~\citep{zang2026sketchtoreality}, and single-sketch mesh generation with shape constraints~\citep{wu2026singleviewsketch} all strengthen single-sketch or sketch-plus-text interpretation.
Model-sheet reconstruction further shows the value of artist-authored multi-view drawings~\citep{yoon2026modelsheets}.
\system differs by deriving multi-view evidence from an in-situ VR sketch, without requiring users to produce polished drawings or model sheets.

\subsection{Sketching Interfaces and VR Spatial Drawing}
\label{sec:related-sketching}

2D sketching interfaces show why drawing is appealing for 3D creation, but also why sketches are challenging inputs for generation.
Teddy~\citep{igarashi1999teddy} creates freeform surfaces from 2D strokes, ILoveSketch~\citep{bae2008ilovesketch} projects 2D strokes to 3D curves for product design, and FiberMesh~\citep{nealen2007fibermesh} uses projected 2D curves as handles for surface modeling.
These systems demonstrate the expressiveness of drawing, but they rely on specialized geometric interpretation or interactive modeling operations rather than treating rough strokes as direct generation conditions.

The literature on drawing-performance reveals that
accurate 2D drawing is challenging
(without tracing)
\citep{Cohen:1997:WCM,Tchalenko:2009:SAC,Schmidt:2009:EPC}.
3D drawing adds additional sources of inaccuracy \cite{arora2017experimental,barrera_machuca_effect_2019,tramper_visuomotor_2011,wiese_investigating_2010},
such as increased motor plan complexity and visual depth perception,
with the benefit of allowing users to directly specify all three dimensions of a 3D curve.
Immersive sketching provides new opportunities for directly-drawn 3D input but amplifies the problem of input inaccuracy.
Lift-Off~\citep{jackson2016liftoff} lifts 2D curves into VR scaffolds, while \citet{arora2017sketching} and Multiplanes~\citep{machuca2018multiplanes} study depth control and assisted freehand drawing.
ScaffoldSketch~\citep{yu2021scaffoldsketch} further targets accurate industrial design drawing in VR through scaffolded drawing support.
Together, these systems show that VR drawing can express spatial intent, but its accuracy depends on depth control, viewpoint management, and the user's motor and drawing skill.
This limitation carries into sketch-conditioned generation: for systems that rely on the user's drawing as the primary shape condition, the final result can be bounded by the user's ability to draw the intended structure accurately in 3D.
Inaccurate, incomplete, or inconsistent strokes can therefore constrain or mislead generation if the sketch is treated as a clean shape specification.
Generative design workflows have also begun to use 3D sketching as an AI-facing structural backbone: Lee et al.~\citep{lee2025carsketching} integrate 3D sketching with 2D generative AI to support view-consistent car exterior renderings.

Recent work connects VR sketching more directly to generated 3D outputs, but still makes different assumptions than we do about the sketch signal.
Luo et al.~\citep{luo2023vrsketch} study VR sketches as a shape-prototyping condition, but not user-defined part labels or part-level text. VRsketch2Gaussian~\citep{gu2025vrsketch2gaussian} combines VR sketches and text for Gaussian object generation, while \system converts labeled strokes into a $2{\times}2$ multi-view part guidance image for textured mesh reconstruction. Order Matters~\citep{chen2025ordermatters} focuses on sequential stroke order as an additional signal, whereas \system treats part channels, part names, and part descriptions as the main structure.
SketchTo3DGen~\citep{shukla2025sketchtogen} is the closest prior system in interaction modality: it combines freehand VR sketches and audio descriptions to generate images and reconstruct 3D assets on a headset.
These works are close to \system because they use 3D drawing as a generative condition, but they do not study the same input setting: rough user-defined part labels, part descriptions, and multi-view guidance for textured mesh generation.
Prior work therefore leaves a specific interface gap: current systems demonstrate the value of spatial drawing, but not how rough, inaccurate, part-labeled 3D strokes can be translated into multi-view guidance for existing open-domain image and 3D generation backbones.
\system targets this gap by treating part-aware VR 3D sketches as approximate spatial guidance, using text descriptions for object and part semantics, and evaluating whether the resulting textured meshes preserve the authored spatial intent.

\section{Design Goals}

People draw for many reasons beyond precise depiction: a rough drawing can externalize an early idea,
mark approximate spatial relationships, and communicate part layout when words are cumbersome.
This includes domains such as characters, architecture, product and vehicle design, scientific illustration, and sculpture.
Text descriptions complement drawing by naming the object and specifying materials, functions, style, and local details that would be difficult or tedious to draw.
At the same time, people typically cannot draw accurately, in 2D or 3D (Section \ref{sec:related-sketching}), because freehand (VR) sketching requires depth control, viewpoint management, and drawing skill.
People often organize and communicate common object concepts in terms of parts and part attributes~\citep{tversky1984objects}. In communicative drawings, people also interpret individual strokes consistently as semantic parts, and strokes depicting the same part tend to cluster during drawing~\citep{mukherjee2019semanticparts}.

We therefore designed \system to translate rough 3D sketches and complementary text descriptions into model-usable conditions for open-domain 3D generation.
(We compare to a version of \system allowing multi-view 2D sketches in our User Study (Section \ref{sec:controlled-user-study}).)
The first three goals articulate this translation problem; the fourth, informed by our pilot study, supports inspection and revision.

\begin{figure*}[t]
  \centering
  \includegraphics[width=\textwidth]{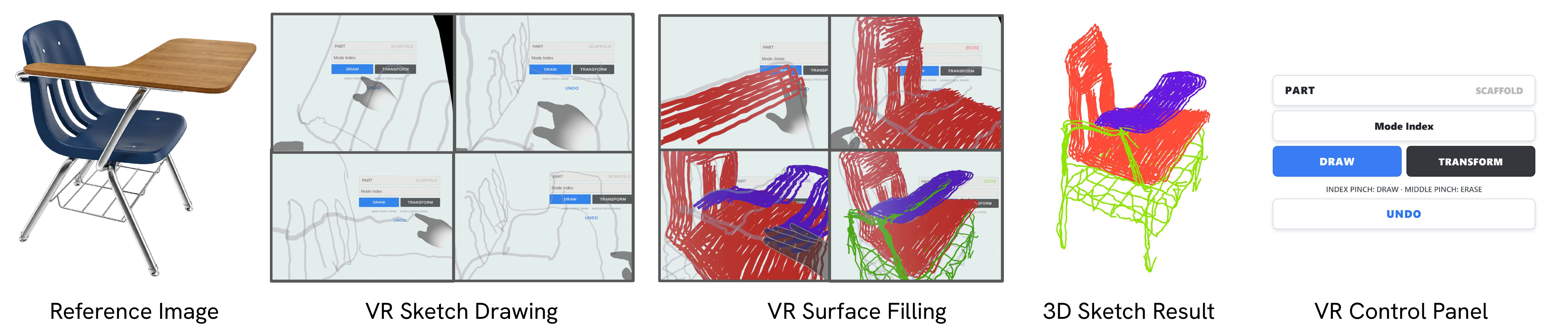}
  \caption{Example VR authoring workflow for a chair with an attached writing tablet. From left to right: the reference image; drawing an unlabeled scaffold in VR; filling in object parts with color-coded strokes; the resulting rough, part-labeled 3D sketch; and the VR control panel for part selection, drawing, transformation, and undo. The scaffold serves as a drawing aid and is excluded from the input used for generation.}
  \Description{Five panels arranged from left to right. A reference photograph shows a chair with an attached writing tablet and an under-seat basket. Four VR screenshots show gray scaffold strokes being drawn, followed by four screenshots showing colored strokes added for object parts. The resulting 3D sketch has red chair surfaces, a purple writing tablet, and a green support frame and basket. The rightmost panel shows the VR controls for part selection, drawing and transformation modes, and undo.}
  \label{fig:vrpipeline}
\end{figure*}

\begin{figure*}[t]
  \centering
  \includegraphics[width=\textwidth]{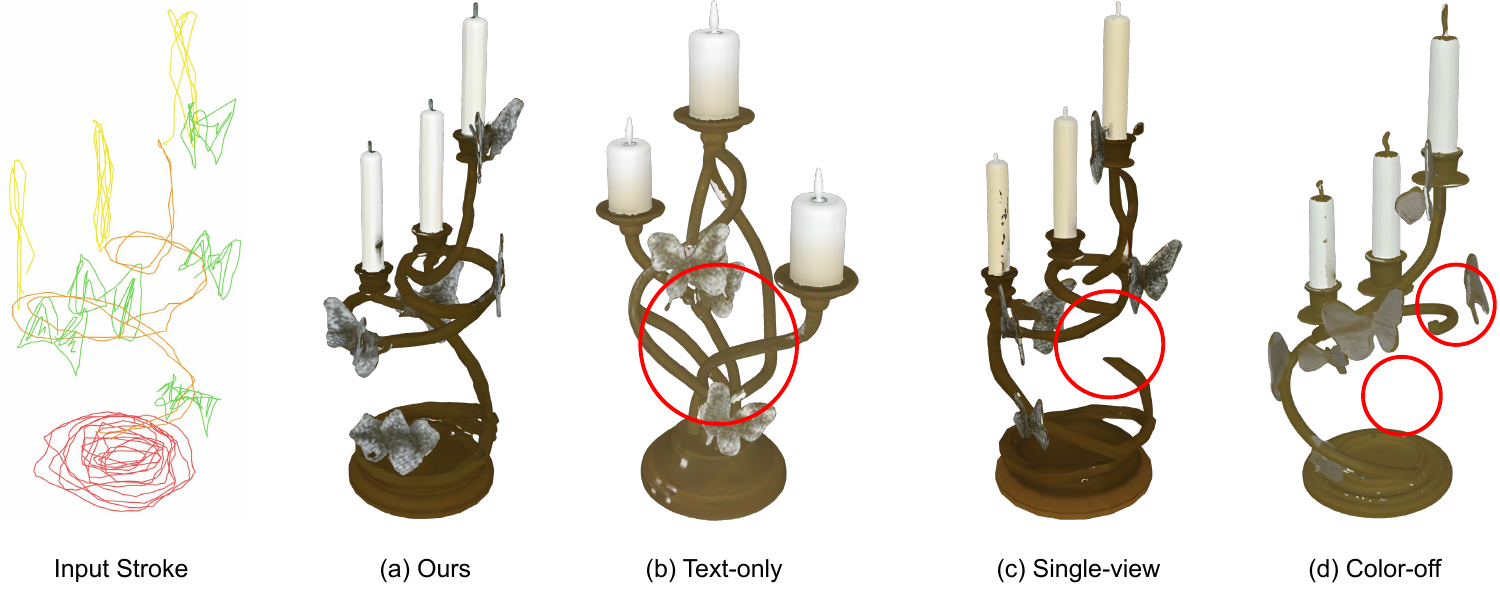}
  \caption{Qualitative comparison motivating spatial guidance and part correspondence. Left: the rough input 3D sketch, with colors indicating parts. (a) Ours preserves the intended stepped arrangement and spiral support. (b) Text-only produces a different branching structure. (c) Single-view reconstruction leaves a gap in the lower support. (d) Color-off omits part of the lower spiral structure and leaves a butterfly detached. Red circles highlight layout discrepancies and incomplete or disconnected structures.}
  \Description{Five panels show a candle-holder sketch and four textured meshes. The left panel shows rough colored strokes for the base, spiral support, candles, and butterfly decorations. Panel (a), Ours, shows three candles at stepped heights on a spiral support. Panel (b), Text-only, shows a different branching support layout, marked with a red circle. Panel (c), Single-view, has a gap in the lower curved support, marked with a red circle. Panel (d), Color-off, has missing lower spiral structure and a detached butterfly on the right, highlighted by two red circles.}
  \label{fig:designgoal}
\end{figure*}

\paragraph{G1: Support imprecise spatial input.}
The system should accept rough strokes made during early ideation, including
strokes that are incomplete, inaccurate, or inconsistent because of drawing
skill or an evolving design idea.
Spatial input portrays shape and spatial relationships that are difficult to capture with text descriptions alone (Figure~\ref{fig:designgoal}(a,b)).
It also provides three-dimensional structural cues that can be lost in single-view reconstruction (Figure~\ref{fig:designgoal}(a,c)).
The 3D sketch should be treated as a rough expression of the intended form, not as precise geometric constraints.
For example, the VR drawing in Figure~\ref{fig:vrpipeline} conveys the relative arrangement of the seat, backrest, writing tablet, and under-seat basket despite uneven strokes and incomplete surfaces.

\paragraph{G2: Combine part-based spatial and text input.}
Rough 3D strokes convey spatial structure, while text specifies object identity, appearance, and local properties that are difficult to draw.
\citet{tversky1984objects} found that people frequently name parts when listing object attributes, and that part names connect appearance with function.
This motivates organizing detailed descriptions around user-defined parts, so users can specify the identity, material, appearance, or function of one component at a time.
The system should therefore let users group strokes into parts and give each group a name and optional description.
Without part grouping, parsing a drawing is often ambiguous.
Figure~\ref{fig:designgoal}(a,d) offers an illustrative comparison: the color-off result has missing support structure and a disconnected ornament.

\paragraph{G3: Open-domain generality.}
The system should support object creation across domains rather than restrict users to a predefined set of categories.
To this end, we aim to leverage the generative capabilities and broad visual priors of existing pretrained image and 3D generative AI models.
The pipeline should translate rough 3D sketches and part descriptions into inputs these models can use, without requiring a new category- or sketch-specific model.
The architectural, product, biological, and character concepts in Figure~\ref{fig:technical-gallery} illustrate this intended breadth.

\paragraph{G4: Support iterative revision.}
Because early design intent often changes after users see concrete outputs, 
revision should be supported as part of the authoring workflow rather than treated as a separate post-processing step.
In our pilot study (Section~\ref{sec:pilot-study}), participants noticed mismatches between their input, intermediate generated images, and the output meshes, but backend reruns required a researcher, motivating user-accessible inspection and revision checkpoints.
The system should let users inspect intermediate images and final meshes so they can decide whether to revise the 3D sketch, 
revise part descriptions, select different views, or rerun a generation stage.

Beyond these design goals, the system should output a standard textured mesh that can be inspected, edited, or imported into downstream 3D workflows.

\section{\system System}
\label{sec:system}

Guided by the design goals above, we developed \system.
\system uses a staged workflow (Figure~\ref{fig:pipeline}).
The pipeline begins with part-labeled 3D strokes from the VR drawing interface, together with object- and part-level descriptions.
\system associates each guidance color with a part name and description and combines these with the optional object-level description.
These inputs condition the generation of candidate multi-view object images.
A selected candidate is then reconstructed as a textured mesh.
The user can revise the drawing input or text descriptions, or rerun generation to explore alternatives.
Our prototype uses two external generation backends: ChatGPT Images 2.0~\citep{openai2025imageapi} for image-conditioned multi-view synthesis and Hunyuan3D-2mv~\citep{hunyuan3d2025} for multi-view image-to-3D reconstruction.
This staged design lets \system translate rough part-labeled VR 3D sketches
and text descriptions into visual conditions that current image-to-3D models can use.

\subsection{Input Representation}
\label{sec:input}

\paragraph{VR 3D drawing.}
\system presents users with a VR drawing interface for drawing, erasing, transforming, and undoing freehand 3D strokes.
Users draw polylines in mid-air with their hand, constructing the object part by part.
Each labeled group of strokes uses a different predefined, visually distinct system color to indicate the part's 3D shape and spatial relation.
Users may also draw unlabeled ``scaffold'' strokes to help align a shape in VR before committing labeled parts; these scaffold strokes are not included in the final 3D sketch passed to the generation stages.
Users operate an in-world control panel (Figure~\ref{fig:vrpipeline}, rightmost panel) with a palm pointer.
From this panel, users select either the unlabeled scaffold or one of up to 20 part slots with distinct, predefined colors; choose among three brushes for small details (index fingertip), medium parts (four fingertips), and broad surfaces (whole hand),
switch between drawing and whole-part transformation, and undo the most recent operation. In drawing mode, pinching the left index finger records strokes and pinching the left middle finger erases near the fingertip. In transformation mode, pinching with both index fingers lets users translate, rotate, and scale the active part.
To revise the input after inspecting a result, users can return to VR, edit the existing strokes with the same drawing and transformation controls, and save the revised sketch for another generation pass.

\paragraph{Preparing generation inputs.}
\label{sec:guidance}

After drawing, users review parts in the GUI, assign names, and provide individual part descriptions.
\system hides scaffold strokes, centers and uniformly scales the remaining sketch, and renders its colored parts from four perspective views at an elevation of $20^\circ$.
Users can select \emph{Choose Front} in the GUI, rotate the sketch preview about the X, Y, or Z axis, and select \emph{Confirm Front} to set the desired front orientation.
The system then regenerates all four guide views relative to this orientation while retaining the fixed camera elevation and field of view.
These views form a $2{\times}2$ guide image on a white background, with front and left views in the top row and back and right views in the bottom row.
The GUI associates each part color with its name and description, then inserts these entries and the object description into a structured prompt.

\subsection{Multi-view Image Synthesis}
\label{sec:prompt}
\label{sec:imagegen}

\system constructs the input to this stage from
the $2{\times}2$ guide image, together with a structured text prompt.
The prompt is assembled from three components: an output specification requesting a clean front, right, back, and left multi-view rendering of a single 
object; a color-to-part mapping that binds each guidance color to its part name; and the user-provided object description, part names, and optional part descriptions.
The full prompt template, output-format requirements, and prompt instructions for the three guide-adherence levels are provided in the supplementary material.

The drawing guidance and structured prompt are passed to a generative image-synthesis backend. The GUI requests multiple candidates at three guide-adherence levels, ranging from stronger preservation of the drawn structure to greater reliance on the text description.
This stage generates a clean $2{\times}2$ multi-view depiction of the target object, because image-to-3D generative models commonly take multiple object views as input.
Each candidate multi-view image contains exactly four views---front, left, back, and right---in the $2{\times}2$ layout described above and provides the visual input for 3D synthesis.

\subsection{Interactive Refinement and 3D Synthesis}
\label{sec:3dgen}

\paragraph{Candidate inspection and revision.}

The GUI displays candidate multi-view image thumbnails and a larger inspection view. Users can select a candidate, request an adjustment through a text description, or generate another set at a different adherence level.
Selecting a candidate uses all four views by default as input to mesh generation. Alternatively, users can choose the front, left, back, and right views from different candidates.
An image can be assigned to a different view than its original label, such as using a candidate's front image as the left view.
Users can leave slots empty and reconstruct from one to four selected views.

\paragraph{3D mesh synthesis.}
The selected object images and their assigned view labels are passed to an image-to-3D reconstruction backend.
The reconstruction model does not accept text prompts; object and part descriptions influence reconstruction through the synthesized multi-view images only.
The 3D synthesis stage converts the generated multi-view depiction into a textured mesh
displayed in the GUI alongside prior reconstructions.

\paragraph{Revision and Editing.}%
\label{sec:revision}
\system supports revision at every stage.
Users can revise the input drawings and text descriptions to refine the intended structure and appearance.
For revising multi-view images, users can drag or paste candidate images, individual views, or external references into the GUI's conversational panel and describe the desired change.
The system combines the edit request and attached images with the original drawing guidance and object and part descriptions to generate revised multi-view candidates.
Users can inspect these candidates and continue the conversation before selecting views for 3D synthesis.
Users can also rerun image generation or mesh synthesis to explore random variations in generative output.
Revisions can address either generative AI mistakes or user preference changes; the latter being a typical part of early-stage ideation.

\subsection{Prototype Implementation}
\label{sec:implementation}

3D sketching takes place in a virtual 3D workspace using hand tracking.
Our implementation runs on a Meta Quest Pro headset.
Input management, image synthesis, candidate adjustment, view composition, and mesh inspection take place in a browser-based GUI outside the headset.

\system renders the 3D drawing into a multi-view guide image using perspective projection with a $48^\circ$ vertical field of view and a $20^\circ$ camera elevation in the coordinate frame defined by the user-selected front orientation, looking toward the sketch center.
(We adopted this elevation from the camera configuration of Instant3dit~\citep{barda2025instant3dit}.)
\system chooses the camera distance that fits the entire sketch in all four views.
Following the prompt's object-centric convention, the semantic views correspond to azimuths of $0^\circ$ (front), $90^\circ$ (left), $180^\circ$ (back), and $270^\circ$ (right).
All views share the same camera distance, field of view, and white background.
\system's multi-view image generation uses the ChatGPT Images 2.0 API \citep{openai2025imageapi}
through its image-editing interface, with the VR guide provided as image input.
In our pilot experiments, this backend performed slightly better than Nano Banana Pro (Gemini 3 Pro Image)~\citep{raisinghani2025nanobananapro} and substantially better than other tested image-generation backends, including an Instant3dit baseline adapted to our test setup~\citep{barda2025instant3dit}.
\system splits each generated $2{\times}2$ image into front, left, back, and right views by locating the central whitespace gutters and removes the white background.
The selected one-to-four images are
passed to the 3D reconstruction model with their user-assigned view labels.
The reconstruction stage uses Hunyuan3D-2mv \citep{hunyuan3d2025}, a feed-forward multi-view image-to-3D model.
We run its multi-view DiT-based shape model to generate an untextured mesh from the selected views, decimate the raw mesh to 100,000 faces using quadric error metrics~\citep{garland1997surface},
and apply Hunyuan3D-Paint \citep{hunyuan3d2025} to produce the final textured mesh (as a glTF file).
Reconstruction runs locally on an NVIDIA H200 GPU server, averaging roughly 17 seconds for shape generation, 28 seconds for texture generation, and 45 seconds in total across the technical-evaluation examples (Section~\ref{sec:technical-evaluation}).

\section{Technical Evaluation}
\label{sec:technical-evaluation}

\begin{figure*}[tb]
  \centering
  \includegraphics[width=\textwidth]{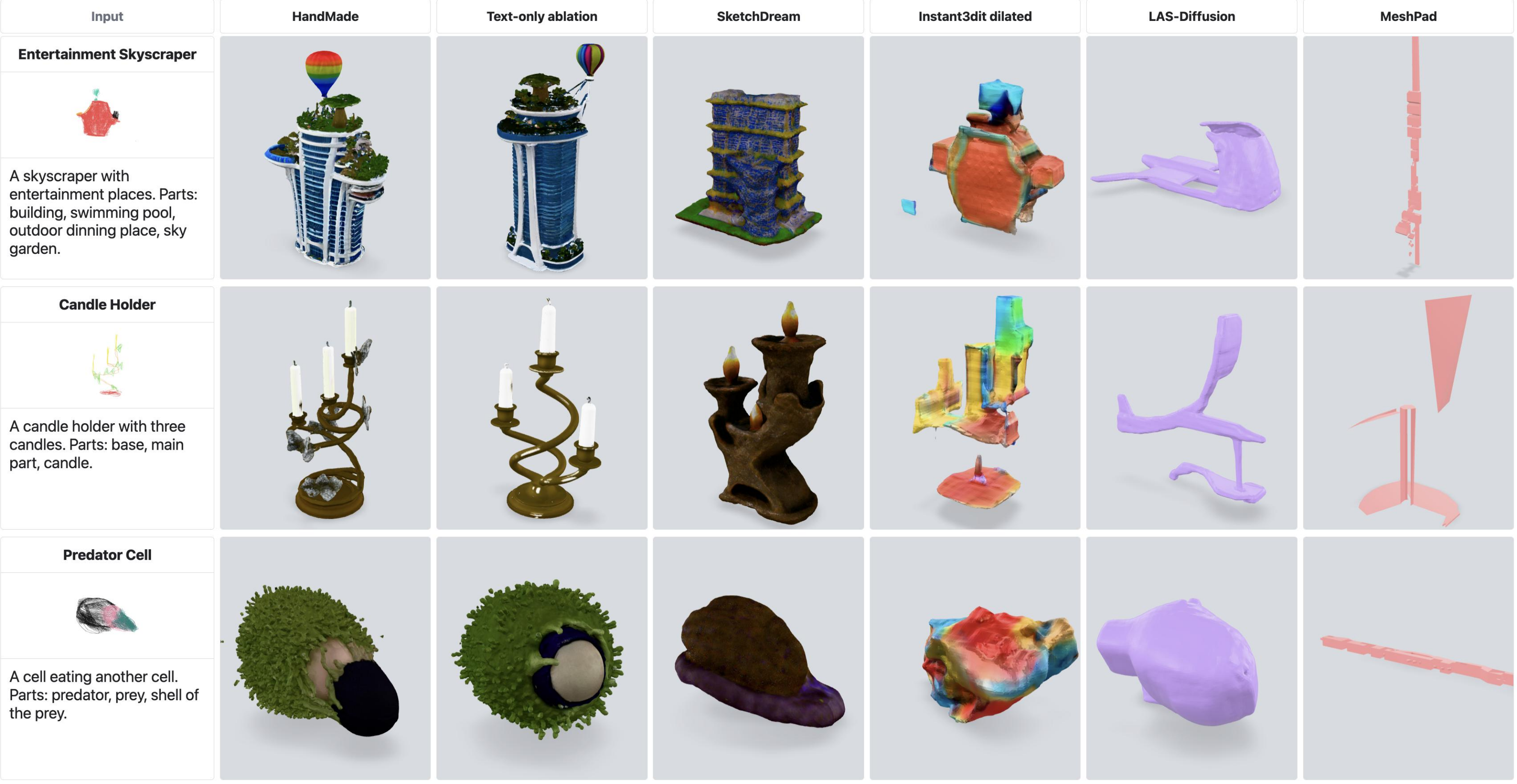}
  \includegraphics[width=\textwidth]{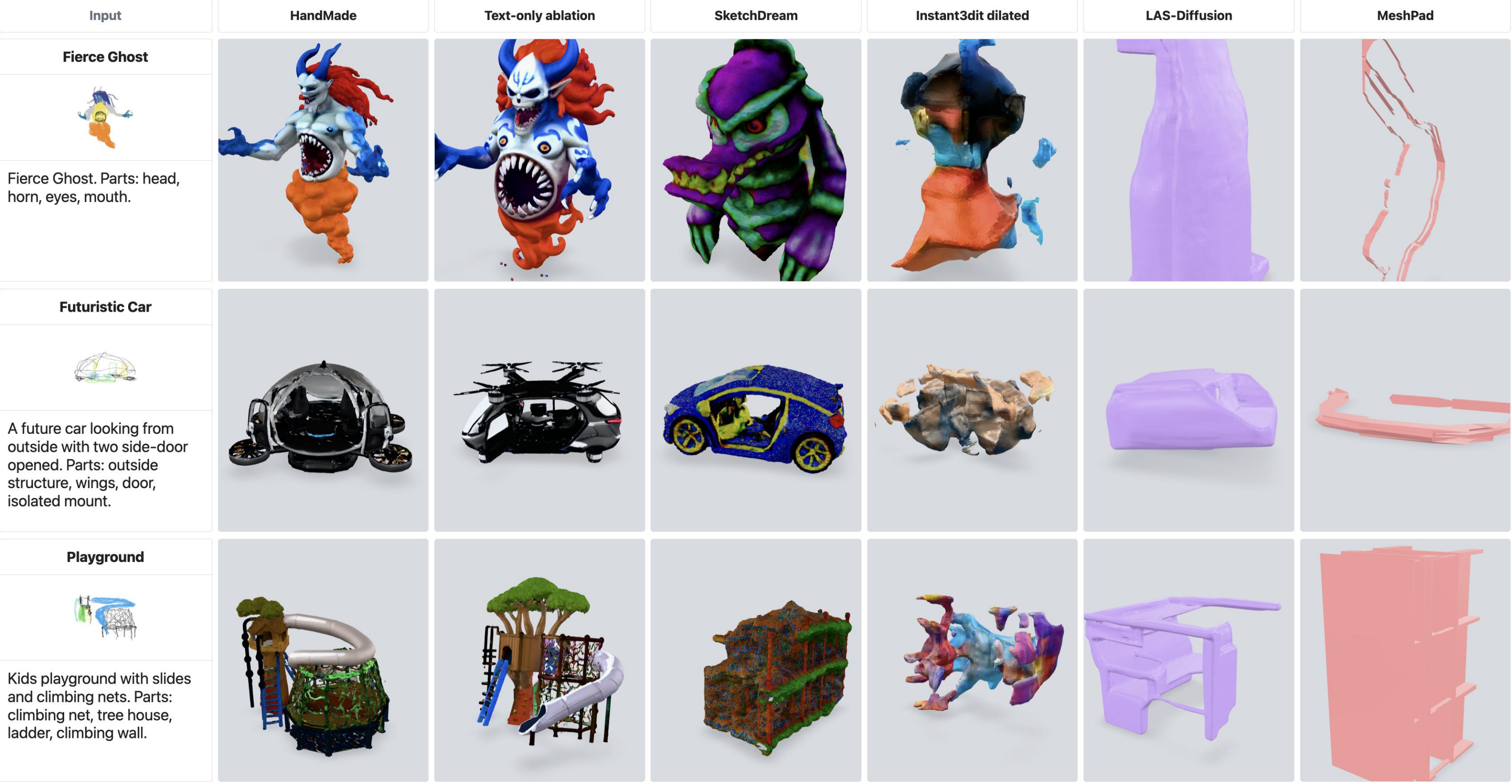}
  \caption{Representative technical-evaluation examples. The gallery compares
  input sketches and generated outputs across \system and baseline conditions,
  emphasizing spatial correspondence, part layout, and semantic completion. The
  text descriptions are simplified here.}
  \label{fig:technical-gallery}
  \Description{Gallery figure comparing input sketches and generated outputs
  across methods.}
\end{figure*}

We compared \textbf{\system}, our full workflow with 3D sketch and text-description input, against two ablations and four baseline methods:
\begin{itemize}
    \item \textbf{Text-only} uses the same user-authored description but omits the rendered 3D sketch guidance from the multi-view image generation stage. This ablation checks the necessity of 3D sketch input.
    \item \textbf{Single-view} uses the same 3D sketch and text descriptions but reconstructs the mesh from a single generated view instead of the multi-view image set. This ablation checks the necessity of multi-view guidance.
    \item \textbf{Instant3dit}~\citep{barda2025instant3dit} is a multi-view inpainting/editing baseline. We project each 3D sketch into Instant3dit's $2{\times}2$ multi-view layout, use a dilated mask around the projected sketch strokes, and reconstruct the edited multi-view image with InstantMesh. This gives Instant3dit the closest baseline input to \system's multi-view spatial guidance while using its own renderer and reconstruction convention.
    \item \textbf{SketchDream}~\citep{liu2024sketchdream} is a sketch-plus-text 3D generation baseline. It requires a sketch image, an RGBA silhouette mask, and a dense depth map. We derive these from the front projection of the normalized 3D sketch rather than using an additional learned depth predictor. This gives SketchDream access to our 3D sketch geometry through its expected input format. In our reproduction, each object requires a separate per-object optimization run, taking roughly 2--3 hours on an 8$\times$H100 node, which makes it substantially more expensive and less interactive than the feed-forward stages in \system.
    \item \textbf{LAS-Diffusion}~\citep{zheng2023lasdiffusion} is a 2D sketch-only ShapeNet/SDF diffusion baseline. We provide the front-view projection of the VR 3D sketch and the required view indicator; the released checkpoint does not use the text description.
    \item \textbf{MeshPad}~\citep{li2025meshpad} is an interactive sketch-conditioned mesh editing system. We adapt its released addition network to single-shot generation by providing the same front-view projection of the 3D sketch from an empty condition; it also receives no text.
\end{itemize}

The evaluation set contains 20 open-domain designs authored as freehand VR 3D sketches with part names and short descriptions.
These designs include the 10 open-domain examples created in the Pilot Study (Section~\ref{sec:pilot-study})
    and 10 others chosen to span a variety of domains:
architecture, vehicles, furniture, playful structures, creature-like forms, and abstract objects. The evaluation focuses on observable geometric preservation: whether generated meshes place surface geometry near the user's drawn 3D structure.
For all methods that accept text, we use the same user-authored description. The comparison therefore holds the high-level authoring intent fixed while adapting the raw 3D sketch to each method's supported input format.

\subsection{Metrics}

We compare each generated mesh with the input 3D sketch in normalized unit-bounding-box space. The 3D sketch is represented by its stroke centerline points, downsampled to at most 8,000 points. Each mesh is represented by 8,000 surface samples. Before scoring, to make sure objects' orientation match, we manually aligned each generated mesh to the 3D sketch in Blender and then applied ICP refinement. We verified that per-point distances are reduced by this alignment.
\begin{itemize}
    \item \emph{Sketch-to-mesh distance} is the mean nearest-neighbor distance from each 3D sketch point to the sampled mesh surface. It measures how well the generated object covers the user's drawn structure.
    \item \emph{Mesh-to-sketch distance} is the mean nearest-neighbor distance from each sampled mesh point to the 3D sketch. It measures how much of the generated surface lies near drawn strokes, but can penalize valid completed surfaces because the 3D sketch is sparse.
    \item \emph{Chamfer distance} is the average of these two one-way mean distances.
    \item \emph{Recall@0.05} is the fraction of sketch points whose nearest mesh-surface distance is below 0.05 normalized units.
    \item \emph{Precision@0.05} is the fraction of mesh surface samples whose nearest sketch distance is below 0.05 normalized units.
    \item \emph{F-score@0.05} is the harmonic mean of Recall@0.05 and Precision@0.05.
\end{itemize}

\subsection{Results}
\label{sec:technical-results}

\begin{table*}[t]
  \centering
  \small
  \setlength{\tabcolsep}{3.2pt}
  \caption{Geometry-alignment results on 20 examples. Distances are computed in normalized unit-bounding-box space using 8,000 sampled mesh-surface points and at most 8,000 sketch points. Recall, precision, and F-score are reported at a threshold of 0.05 normalized units. The best score is in bold.}
  \label{tab:technical-metrics}
  \begin{tabular}{lcccccc}
    \toprule
    Method & Sketch$\to$Mesh $\downarrow$ & Mesh$\to$Sketch $\downarrow$ & Chamfer $\downarrow$ & Recall@0.05 $\uparrow$ & Precision@0.05 $\uparrow$ & F-score@0.05 $\uparrow$ \\
    \midrule
    \system & \textbf{0.035} & \textbf{0.043} & \textbf{0.039} & \textbf{0.792} & \textbf{0.685} & \textbf{0.727} \\
    Single-view & 0.040 & 0.048 & 0.044 & 0.719 & 0.642 & 0.670 \\
    Text-only & 0.047 & 0.058 & 0.052 & 0.652 & 0.539 & 0.583 \\
    Instant3dit & 0.037 & 0.045 & 0.041 & 0.748 & 0.658 & 0.693 \\
    SketchDream & 0.041 & 0.064 & 0.053 & 0.716 & 0.497 & 0.576 \\
    LAS-Diffusion & 0.058 & 0.073 & 0.065 & 0.581 & 0.457 & 0.501 \\
    MeshPad & 0.096 & 0.080 & 0.088 & 0.340 & 0.456 & 0.361 \\
    \bottomrule
  \end{tabular}
\end{table*}

Across the 20 examples, \system achieves the strongest average geometry alignment among the alternatives (Single-view, Text-only, and baseline algorithms)
(Table~\ref{tab:technical-metrics}). This supports the central technical claim that part-labeled VR sketches provide useful spatial information for open-domain generation. We interpret the numbers as evidence of sketch-geometry preservation, not as a broad claim of overall asset quality.

The Text-only ablation often produces plausible objects, which is expected from a strong generative prior and a descriptive prompt. However, without the sketch it frequently misses the specific geometric correspondence that matters for an authored design: the size, placement, and orientation of parts. \system improves Recall@0.05 over Text-only on 16 of the 20 examples. The largest gains appear on objects whose layouts are hard to specify in text alone, such as \emph{Stacked Sphere Table}, \emph{Dragon Ship}, \emph{Towel Hanger}, \emph{Predator Cell}, and \emph{Candle Holder}. In these examples, text names the object and its semantic attributes, while the sketch supplies the spatial arrangement that makes the design specific.

The Single-view ablation isolates the value of multi-view reconstruction guidance. It performs better than Text-only on average, showing that even a single visual condition carries useful spatial information. However, \system improves Recall@0.05 over Single-view on 17 of the 20 examples, with especially large gains on \emph{Towel Hanger}, \emph{Playground}, \emph{Planet Model}, \emph{Chandelier}, \emph{Candle Holder}, and \emph{Futuristic Car}. These cases contain cross-view structure, elongated parts, or part relationships that are ambiguous from one projection. The multi-view guidance therefore helps the reconstruction stage preserve 3D arrangements that are not fully visible from a single image.

Instant3dit is the strongest baseline and the closest comparison to \system's input setting. Its aggregate metrics are near \system's.
This indicates that multi-view projected sketches are a useful input to existing inpainting/editing pipelines. However, visual inspection (Figure~\ref{fig:technical-gallery}) shows that Instant3dit reinterprets unusual layouts, weakens part relationships, and produces objects that align with the strokes geometrically while differing in object structure or semantics.

SketchDream is also a meaningful comparison because it accepts both sketch and text. It often recovers broad object identity and is strong on some familiar or compact shapes. Its average Recall@0.05 is close to Single-view, but its lower Precision@0.05 and F-score@0.05 indicate that its optimized shapes often add surface geometry that is farther from the sparse sketch. Combined with the per-object optimization cost described above, this makes SketchDream useful as a quality-oriented baseline albeit less suitable for the interactive workflow targeted by \system.

LAS-Diffusion and MeshPad are more strongly constrained by their training domains and input assumptions. LAS-Diffusion sometimes obtains reasonable geometric overlap on shapes that resemble its ShapeNet-style prior, but without text it often produces smooth category-prior surfaces that do not match the intended object semantics. MeshPad is even more mismatched for this task: its single-shot adaptation often produces small fragments or patch-like meshes rather than complete open-domain objects. These results clarify why front-view 2D sketch-to-shape comparisons are insufficient for our setting. Much of the user's intent is expressed in 3D part layout and cross-view structure, not in a single silhouette.

Overall, the technical evaluation demonstrates the value of multi-view spatial guidance paired with text descriptions, and the insufficiency of baseline approaches when the user's design deviates from category priors.

\section{User Evaluation}
\label{sec:user-study}

We evaluated \system in two stages. A 10-participant pilot study examined how people used the initial VR-centered workflow and informed the interface redesign. We then conducted an 18-participant within-subject study comparing VR 3D drawing and multi-view 2D drawing while holding the downstream GUI-based generation pipeline and target object constant. Our studies were approved by our institutional ethics board.

\subsection{Pilot Study}
\label{sec:pilot-study}
Ten participants (P1--P10) used a prototype, VR-centered version of \system to create an object of their own imagination. In this version, participants used a physical keyboard visible through VR passthrough~\citep{meta2026passthrough} to start and stop drawing, switch part channels and modes, enter object and part descriptions, and undo operations. After authoring, a researcher manually ran the multi-view image-generation and 3D synthesis stages and showed participants the generated multi-view images and resulting meshes. Revision options were to edit the VR strokes or text, or ask the researcher to rerun a backend stage. Two preliminary sessions (P1--P2) also included a guided practice object, which we removed because it lengthened training and appeared to prime participants' object choices. The examples across all ten sessions spanned characters, architecture, products and vehicles, scientific illustrations, and sculpture. We collected session observations, open-ended feedback, and questionnaire responses. We report the qualitative findings here and descriptive questionnaire results for P3--P10 in the supplementary material, excluding the two preliminary sessions from the quantitative summary.

Operating the physical keyboard through passthrough was cumbersome. P3 reported that ``using the keyboard to switch was hard,'' and P10 often forgot shortcuts.
P6 and P8 reported dizziness, with P8 writing, ``The view makes me so dizzy.''
These observations motivated replacing the passthrough-and-keyboard workflow with in-world controls and pinch gestures in a fully virtual workspace.

Participants also wanted to manipulate the drawing itself. P5 asked to ``rotate so I don't need to walk around.''
We added direct translation, rotation, and scaling of the active part so users can adjust its placement and inspect it without relying only on their own movement.
Reported difficulty with small and thin features also motivated brush configurations for different drawing scales.

Revision depended on researcher-run backend stages, limiting participants' direct control. Session feedback called for more freedom to revise generated results.
P4 wanted the system to ``ask back about the details of the shape,'' while P3 was satisfied with the multi-view image but not the resulting mesh.
These observations motivated the GUI (Figure~\ref{fig:gui}) described in Section~\ref{sec:implementation}, which lets users edit descriptions, inspect candidates, request image adjustments, select views, and rerun 3D synthesis directly.

\subsection{User Study}
\label{sec:controlled-user-study}
Following the pilot study, we created a non-VR GUI and updated the VR drawing interactions based on participants' feedback.
We then conducted a controlled user study to understand how participants used the revised drawing-and-generation workflow and how satisfied they were with the results.
To examine the value of drawing directly in 3D, we also compared it with multi-view 2D drawing, an alternative input to the same generation pipeline.

\subsubsection{Research Questions}

The study addressed three questions:
\begin{itemize}
  \item \textbf{RQ1 (workflow experience):} How do participants experience the workflow, including workload, revision, and satisfaction with the generated results?
  \item \textbf{RQ2 (spatial expression):} How does 3D drawing compare with multi-view 2D drawing for expressing and correcting spatial structure?
  \item \textbf{RQ3 (preference):} Which input method would participants choose for a similar future task, and why?
\end{itemize}

\subsubsection{Participants, Design, and Procedure}

Eighteen participants (U1--U18) completed the study (ages 24--39 years, mean 29.1, standard deviation 3.8).
Most had limited experience with 3D authoring: 14 rated their CAD, 3D modeling, or sculpting experience at 1 or 2 on a 5-point scale, and 14 gave the same ratings for creating or editing content in VR/AR.
Three participants rated their 2D drawing experience at 4 or 5.

We used a within-subject design with counterbalanced order. Nine participants completed the 3D condition first and nine completed the 2D condition first.
In both conditions, participants reproduced the same school chair with an attached writing desk and under-seat basket (Figure \ref{fig:vrpipeline}).
The target combined familiar chair structure with asymmetric and partially occluded parts, requiring participants to convey their relative positions and connections.
At the start of each workflow, a researcher guided participants in observing the target as a virtual 3D model.
The target was then hidden for the remainder of that workflow, so participants drew and evaluated their results from memory.

In the \textbf{3D condition}, participants drew the target in VR using color-coded parts.
In the \textbf{2D condition}, they drew one or more views on an iPad, using the same color for the same part across views.
Participants kept the same part grouping across conditions: for example, if they grouped the basket and chair legs as one part in the first workflow, they kept that grouping in the second.
They then continued in the GUI, where we added a 2D input option for uploading drawings and assigning viewpoint labels.
From this point, both conditions followed the same process: participants typed in the part names and descriptions, generated multi-view image candidates, optionally revised or regenerated them, selected views for 3D synthesis, and inspected the final mesh.
For the second workflow, they could copy the text descriptions from the first.

For the 2D condition, we adapted the multi-view image generation prompt template to the independently created, viewpoint-labeled drawings and asked the model to reconcile cross-view inconsistencies.
Color-to-part mappings and the four-view output layout were retained.
The full initial-generation prompt templates and guide-adherence policies for both conditions are provided in the supplementary material.

Instruction and guidance averaged 11.6 minutes.
Participants were asked to complete each workflow in approximately 20 minutes, although we retained completion times for sessions that ran longer.
Participants completed the corresponding questionnaire immediately after each workflow.
After both workflows, they answered the direct comparison questions and provided open-ended feedback.

\subsubsection{Measures and Analysis}
\label{sec:user-study-measures}

Participants rated each workflow on 15 five-point Likert scale questions covering their recollection of the target, spatial expression, ease of correction, perceived control, result quality and satisfaction, revision support, and workload.
Four direct comparison questions asked which workflow made spatial expression and correction easier, produced a better match to the target, and would be preferred for a similar future task.
Each allowed an ``about the same'' response.
We also recorded drawing, generation/revision, and total workflow time.

\begin{figure*}[t]
    \centering
    \includegraphics[width=\linewidth]{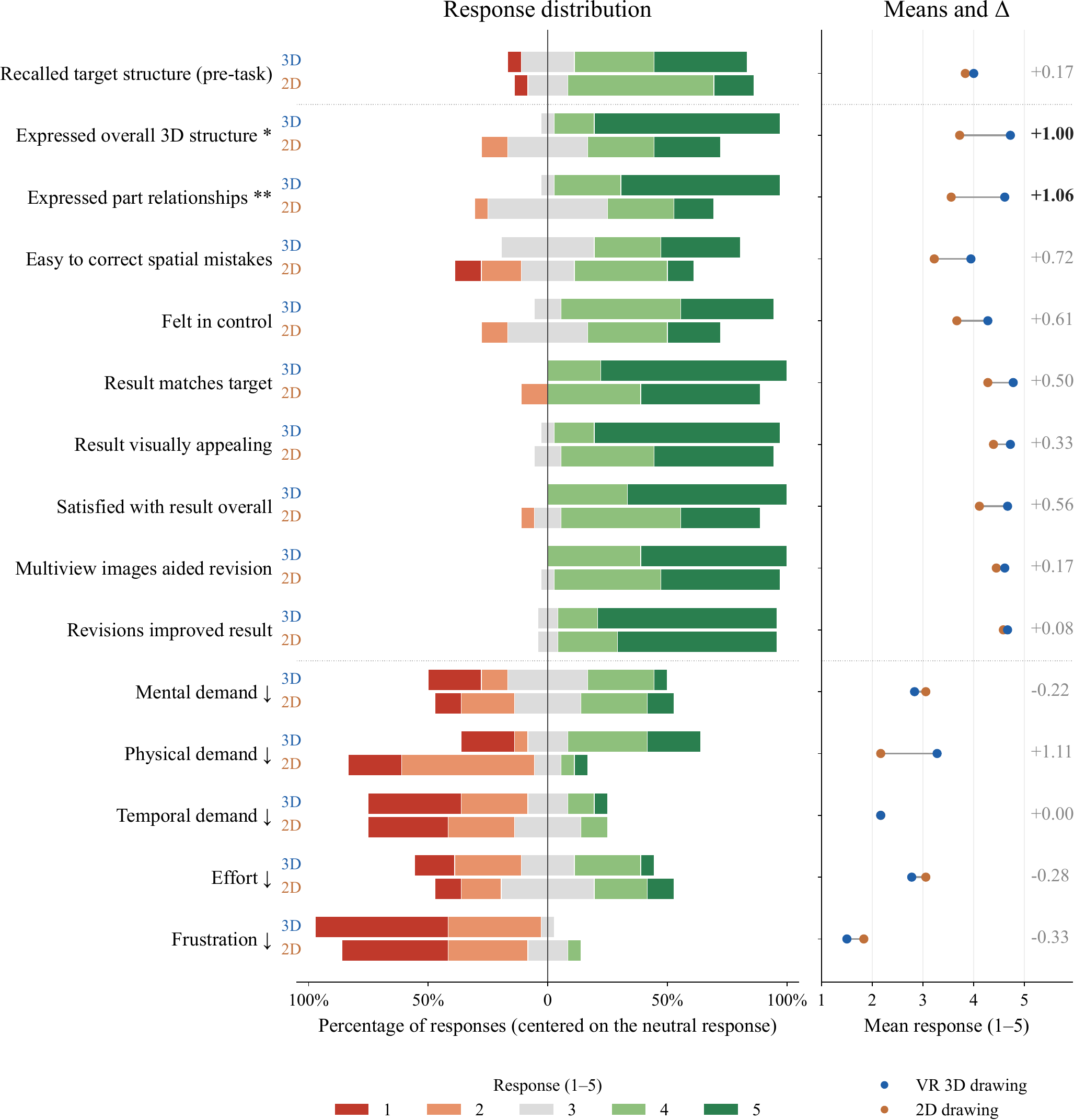}
    \caption{A visualization of paired questionnaire results from Table \ref{tab:user-study-likert}.}
    \Description{A horizontal bar chart visualization of the data in Table \ref{tab:user-study-likert}.}
    \label{fig:user-study-likert}
\end{figure*}

We analyzed paired ratings with two-sided Wilcoxon signed-rank tests, computing exact $p$ values by sign randomization with zero differences omitted and average ranks assigned to ties.
We applied Holm-Bonferroni correction across all 15 items.
Table~\ref{tab:user-study-likert} reports condition means, paired mean differences, and adjusted $p$ values for Q1--Q15.
Figure~\ref{fig:user-study-likert} visualizes this data.
Each comparison uses complete pairs; the revision-improvement item has 12.
Direct choices were analyzed with exact two-sided binomial tests excluding ties and Holm-Bonferroni correction across the four questions.
We corrected timing comparisons across the three measures and examined condition order and period because participants used the same target and GUI twice.
Open-ended responses helped explain participants' choices; comparisons of preference with prior experience or satisfaction are descriptive.

We also compared each participant's final mesh from each condition with the ground-truth (GT) chair model, giving 18 paired comparisons.
We sampled 8,000 points from each surface in proportion to area.
Registration used a similarity transform that optimized rotation, a single uniform scale, and translation.
With the GT oriented to match the outputs, we initialized alignment from the meshes' bounding boxes and used bidirectional iterative closest point registration with the closest 90\% of correspondences, with up to 15 initial and 50 refinement iterations.
After alignment, we measured the
normalized Chamfer distance (average Euclidean distance from every sample point to its nearest sample point on the other mesh, averaged in both directions, divided by the bounding-box diagonal).
Scoring used all sample points, without trimming.
Lower values indicate closer surface agreement.
We then calculated the mean and standard deviation across the 18 final meshes in each condition.
We used a separate exploratory two-sided exact paired signed-rank test for this geometric comparison.

\subsection{User Study Results}
\label{sec:user-study-results}

\subsubsection{Comparison with Ground Truth}
\label{sec:user-study-gt}

Meshes produced with 3D drawing were closer to the ground-truth chair on average.
We compared the aligned mesh surfaces using the normalized Chamfer distance described in Section~\ref{sec:user-study-measures}.
The mean normalized Chamfer distance was 0.02381 (standard deviation 0.00366) for 3D drawing and 0.02597 (standard deviation 0.00420) for 2D drawing, an 8.3\% reduction.
The 3D result had lower Chamfer distance in 15 of the 18 participant pairs (two-sided exact paired signed-rank test: $W=34$, $p=.024$).
This comparison measures how closely the completed workflows reproduced the shared chair target after participants drew from memory.

\begin{table*}[t]
  \centering
  \caption{Paired questionnaire results. $n$ gives the number of complete pairs, $M$ the condition mean, and $\Delta$ the mean 3D-minus-2D difference. Bold adjusted $p_{\mathrm{H}}$ values indicate significant paired differences ($p_{\mathrm{H}}<.05$). $\downarrow$ marks workload items, for which lower ratings are favorable. See Figure \ref{fig:user-study-likert} for a visualization.}
  \label{tab:user-study-likert}
  \small
\begin{tabular}{lrrrrr}
\toprule
Question & $n$ & 3D $M$ & 2D $M$ & $\Delta$ & $p_{\mathrm{H}}$ \\
\midrule
Q1. Recalled target structure (pre-task) & 18 & 4.00 & 3.83 & +0.17 & 1.000 \\
\midrule
Q2. Expressed overall 3D structure & 18 & 4.72 & 3.72 & +1.00 & \textbf{0.014} \\
Q3. Expressed part relationships & 18 & 4.61 & 3.56 & +1.06 & \textbf{0.009} \\
Q4. Easy to correct spatial mistakes & 18 & 3.94 & 3.22 & +0.72 & 0.521 \\
Q5. Felt in control & 18 & 4.28 & 3.67 & +0.61 & 0.605 \\
Q6. Result matches target & 18 & 4.78 & 4.28 & +0.50 & 0.668 \\
Q7. Result visually appealing & 18 & 4.72 & 4.39 & +0.33 & 1.000 \\
Q8. Satisfied with result overall & 18 & 4.67 & 4.11 & +0.56 & 0.102 \\
Q9. Multi-view images aided revision & 18 & 4.61 & 4.44 & +0.17 & 1.000 \\
Q10. Revisions improved result & 12 & 4.67 & 4.58 & +0.08 & 1.000 \\
\midrule
Q11. Mental demand$\downarrow$ & 18 & 2.83 & 3.06 & -0.22 & 1.000 \\
Q12. Physical demand$\downarrow$ & 18 & 3.28 & 2.17 & +1.11 & 0.309 \\
Q13. Temporal demand$\downarrow$ & 18 & 2.17 & 2.17 & +0.00 & 1.000 \\
Q14. Effort$\downarrow$ & 18 & 2.78 & 3.06 & -0.28 & 1.000 \\
Q15. Frustration$\downarrow$ & 18 & 1.50 & 1.83 & -0.33 & 1.000 \\
\bottomrule
\end{tabular}

\end{table*}

\subsubsection{Spatial Expression and Coherence}
The clearest rating differences concerned expressing spatial structure.
Participants rated 3D significantly higher for overall structure (Q2: 3D $M=4.72$, 2D $M=3.72$, $p_{\mathrm{H}}=.014$) and relationships among parts (Q3: 3D $M=4.61$, 2D $M=3.56$, $p_{\mathrm{H}}=.009$).
In direct comparisons, 16 chose 3D for easier spatial expression, one chose 2D, and one chose about the same ($p_{\mathrm{H}}=.0011$).

Participants described the advantage as keeping the object consistent across views.
3D drawing is multi-view consistent \emph{by construction}, whereas multi-view 2D drawings require great care for consistency.
U11 omitted the writing tablet's L shape in 2D because they had not thought to include a top view; in VR, they felt they were editing ``one single coherent object.''
U10, with over six years of professional sketching training, found separate 2D perspectives difficult to align and the whole model easier to envision in 3D.

\subsubsection{Revision}
Participants valued the intermediate multi-view images as a checkpoint before finalizing their results.
They rated the images as helpful for deciding whether and how to revise in both conditions (Q9: 3D $M=4.61$, 2D $M=4.44$).
Among the 12 participants with paired revision ratings, perceived improvement toward the target also received high ratings (Q10: 3D $M=4.67$, 2D $M=4.58$).

When asked which workflow made spatial mistakes easier to correct, 13 participants chose 3D, two chose 2D, and three chose about the same ($p_{\mathrm{H}}=.0148$).
U12, with over twenty years of amateur drawing experience, found spatial mistakes easier to correct in VR despite its greater physical demands.
Some drawing operations remained difficult: U5 found colored parts harder to erase and revise than the scaffold, while U14 struggled to draw flat surfaces in 3D.

Across six categories of revision actions---VR drawing updates, 2D drawing updates, text-prompt changes, image regeneration, conversational editing, and mesh regeneration---we recorded 23 actions in 3D and 30 in 2D, averaging 1.28 and 1.67 per session, respectively.
Initial input preparation and first image and mesh generations were excluded.
Image regeneration occurred seven times in each condition, while additional mesh synthesis occurred nine times in 3D and ten in 2D.

Inspection of the 12 conversational requests identified changes to geometry, part relationships, or cross-view consistency; none requested appearance or style changes.
In five sessions, participants combined AI-adjusted views with unedited candidate views as 3D synthesis inputs.
Researcher observation also documented U12 returning to VR to remove a frame segment that was difficult to specify verbally.

\subsubsection{Output Quality}
Participants rated the final outputs positively in both conditions.
Mean ratings for target match (Q6) and visual appeal (Q7) exceeded 4 in both conditions, and overall satisfaction (Q8) averaged 4.67 for 3D and 4.11 for 2D (Table~\ref{tab:user-study-likert}).
All 18 participants rated their 3D result at 4 or 5 for both target match and satisfaction; in 2D, 16 gave these ratings for target match and 15 for satisfaction.
In the direct target-match comparison, 11 participants chose the 3D result, seven chose about the same, and none chose 2D ($p_{\mathrm{H}}=.0029$).

\subsubsection{Future Preference and Trade-offs}
Fifteen participants would choose 3D for a similar future task, while three would choose 2D ($p_{\mathrm{H}}=.0148$). This included all three participants with 2D drawing experience ratings of 4 or 5, and five of the six who gave their 2D result the highest satisfaction rating.

For U15, two 2D views were enough to obtain what they considered a similar-quality result.
They preferred 2D for this task, suggesting that 3D drawing might be more worthwhile for complex objects or scenes.
The other two participants who chose 2D raised practical concerns: U2 needed time to become comfortable in VR despite finding the results satisfying, and U6 questioned buying a VR device solely for object design.

Both workflows took about 20 minutes on average (3D: 21.67 minutes; 2D: 20.11 minutes).
Completion time also depended on condition order: the 3D workflow took 7.89 minutes longer in the 3D-first group but 4.78 minutes less in the 2D-first group (order-label permutation $p_{\mathrm{H}}=.002$).
The second workflow was 6.33 minutes faster on average, consistent with familiarity with the task and GUI and the opportunity to reuse descriptions.

\clearpage
\section{Discussion}

\system shows that rough 3D sketches can function as spatial prompts for
open-domain 3D generative AI models.
Namely,
\system demonstrates a translation workflow between early human spatial
intent and the input formats that generative models can use. Within this
workflow,
sparse 3D strokes externalize where parts should be, how they relate, and
what spatial organization makes an object specific, while text descriptions supply
identity, material, style, and local semantic detail.
This combination of 3D sketching, text descriptions, staged generation,
and revision
lets users move from an imagined
spatial arrangement to a textured 3D asset.

\subsection{Spatial Prompting}

Our evaluations highlight the value of directly expressing 3D shape and spatial relationships.
The technical comparison found closer geometric alignment with the input sketches than text-only generation (Section~\ref{sec:technical-results}).
In the controlled study, participants rated VR drawing more highly for expressing overall structure and part relationships, and the final meshes had lower Chamfer distance to the ground-truth chair (Section~\ref{sec:controlled-user-study}).
U10 and U11's accounts illustrate the difficulty of coordinating separate 2D perspectives, whereas drawing in 3D provides a shared spatial representation from which multiple views can be derived.

Preference for the 3D workflow also included all three participants who rated their 2D drawing experience at 4 or 5.
Within this sample, preference for direct 3D authoring was therefore not confined to participants with limited 2D drawing experience.
This observation supports considering rough 3D drawing not only as an entry point for novices, but also as a potential complement to existing drawing skills.

Prior sketch-based modeling and immersive sketching systems show the expressive
value of drawing for shape ideation and spatial authoring. \system builds on
this tradition, but expands the role of the sketch;
a part-labeled 3D sketch becomes a rough generative guide that carries approximate part presence, extent, orientation, and spatial relations into a generative workflow,
while being robust to varying degrees of inaccuracy.

\subsection{Legible Intermediate Representations}

Prompting research has shown that generative AI workflows often shift effort from manual production to prompt formulation, inspection, and iteration~\citep{tankelevitch2024metacognitive,weisz2024designprinciples}.
In \system, the useful intermediate representation is the synthesized multi-view image.
Its value extends beyond making generated results inspectable. It lets users determine which generated content becomes input to the next stage.
In five sessions, participants combined AI-adjusted views with unedited candidate views for 3D synthesis.
These actions separate the generation of a candidate from acceptance of its complete view set, allowing existing image inputs to be carried forward alongside revisions.

This distinction matters because completing an under-specified drawing requires the generative model to make choices that the user has not explicitly authored.
An editable intermediate representation lets the user intervene in which of those generated proposals are passed forward, rather than accepting or rejecting an entire candidate.
The checkpoint therefore supports selective reuse of generated content, not only repeated generation.

These intermediate representations let the generative stages work with approximate spatial guidance and imperfect multi-view images.
The input 3D sketch and descriptions may be rough, incomplete, or inconsistent, yet the rendered guide can still preserve part-level spatial evidence that multi-view image synthesis can use.
The generated multi-view images may also contain view inconsistencies, yet 3D shape synthesis can still recover a coherent mesh when the views preserve enough shared object structure.
Thus, the pipeline works by translating imperfect evidence into progressively more model-usable representations, without assuming that either the human input or the generated images are already precise.

The current GUI lets users enable/disable input parts, edit descriptions, adjust guide adherence, revise image candidates, and compose selected views.
Future interfaces should extend this control to
masking and regenerating only parts of image candidates or the mesh surface.
In this sense, the intermediate representations
are not merely implementation details; they are interaction objects that help
users maintain agency over a generative pipeline.

\subsection{Global Versus Spatially Grounded Revision}

\begin{figure*}[t]
  \centering
  \includegraphics[width=\textwidth]{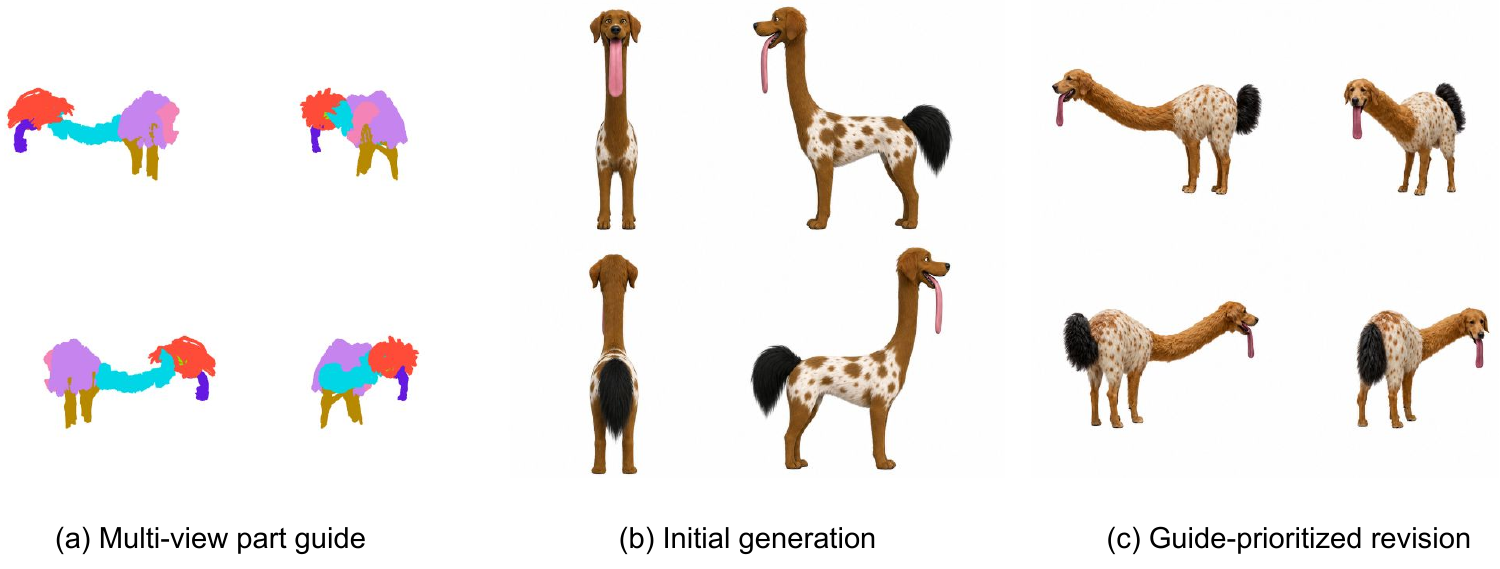}
  \caption{Qualitative example of guide adherence in the multi-view image synthesis stage. (a) A multi-view part guide rendered from the VR sketch. (b) An initial generation that is semantically plausible but deviates from the elongated spatial guide. (c) A guide-prioritized revision that more closely preserves the intended structure. \system's GUI exposes this tradeoff through three guide-adherence levels.
  }
  \label{fig:spatial-guidance-case}
  \Description{Three side-by-side panels showing (a) a multi-view part guide, (b) an initially generated dog with a more typical body shape, and (c) a guide-prioritized revision with a more elongated body matching the guide.}
\end{figure*}

\system's GUI exposes three guide-adherence levels, allowing users to specify how strictly image candidates
should follow the drawn structure versus a more permissive interpretation
favoring the descriptions and the model prior.
Figure~\ref{fig:spatial-guidance-case} compares the multi-view part guide (a), the initial generation (b), and a guide-prioritized revision (c). The initial generation
deviated from the elongated guide;
a revision prioritizing the guide preserves the intended structure more closely.
However, guidance strength and the spatial target of an edit are distinct aspects of control. Following a drawing more closely does not itself specify which local region or relationship a revision should change.

The structural content of the 12 conversational requests shows that language can express geometric and relational changes.
Researcher observation documented U12 returning to VR to remove a small section of the chair’s support frame that was difficult to identify verbally. This case illustrates a potential gap between describing a change and identifying the geometry to be changed.
U12’s case suggests that direct spatial selection could complement verbal instructions when users know what change they want but have difficulty identifying the intended part or region in words.
Although \system associates descriptions with named parts, local relational instructions lack explicit spatial grounding.
The interface does not provide direct selection of the local stroke region, contact area, or surface patch to which an instruction applies.
Future interfaces could combine adjustable adherence with spatially grounded instructions, letting users select the parts or regions involved and describe the intended relationship.
This would extend control from globally following a drawing to specifying where and how generation should respond.

\section{Limitations and Future Work}

\system treats rough 3D sketches as approximate spatial guides: useful for expressing approximate part-level forms and spatial relations, but incomplete for surface detail, hidden geometry, topology, and exact
dimensions. The input itself is also limited evidence of intent: rough VR
strokes can reflect hand-tracking noise, depth-control difficulty, drawing skill,
or an idea that changes once the user sees generated candidates. As a result,
the evaluation can measure preservation of the authored drawing, but not perfect
faithfulness to the user's unobserved mental model.

Failure modes arise at several stages of the pipeline. At the input stage, a sketch can under-specify contact, symmetry, flatness, or which nearby parts should
remain separate.
Rendering the 3D sketch into a fixed set of multi-view guidance images enables compatibility with existing image-generation models, but does not convey all of the authored 3D information.
Occluded geometry and spatial relationships that become ambiguous in projection may be lost or weakened before generation.
At the multi-view image synthesis stage, the image generator can ignore a
guidance color, reinterpret an unusual proportion through a familiar category
prior, alter part scale, or introduce multi-view inconsistencies. At the
3D shape synthesis stage, the mesh generator can smooth thin structures, merge adjacent
parts, lose holes or hidden geometry, or convert a plausible image into a mesh
whose topology is difficult to edit.

Our baseline comparisons are limited to available implementations that can be adapted to our input setting.
Geometric alignment measures proximity to the input sketch or target surface, but does not establish semantic part correspondence, structural completeness, or correct connectivity.

We measured perceived target match, visual appeal, and overall satisfaction separately (Q6--Q8), but did not establish how the first two contributed to satisfaction.
In open-ended creation, satisfaction need not imply fidelity to an initial idea.
Future studies should track evolving intentions and assess intent match and visual appeal separately.

The user study also identified an interaction tradeoff: VR 3D drawing supported coherent-by-construction spatial expression but was physically more demanding on average than multi-view 2D drawing. Participants continued to report difficulty drawing flat surfaces and managing large or occluded structures in mid-air.
Future work could close the loop by comparing generated meshes with the original VR sketch.
After alignment, geometric deviations could be localized along the strokes and aggregated by user-defined parts to guide targeted revision.
Because sketches are approximate and incomplete, the loop should not minimize all differences indiscriminately; users could confirm which deviations warrant correction before regeneration.

\section{Conclusion}

We presented \system, a system that
treats rough, part-labeled 3D strokes as spatial guides and translates them into editable multi-view conditions for modern image and 3D generative AI models.
By combining 3D sketching with text descriptions, \system gives users a more direct way to communicate object layout than text prompting alone, while still benefiting from the broad priors of open-domain generative models.
Our evaluations support the value of rough 3D sketches as spatial guidance for generation.
Participants used the end-to-end workflow to create and revise 3D assets, rated the generated results positively, and generally preferred VR drawing for similar future tasks.
VR drawing also received higher spatial-expression ratings and yielded meshes closer to the target chair after alignment, although physical-demand ratings were higher on average.
Multi-view image synthesis can turn noisy or incomplete drawings and text descriptions into more complete visual depictions suitable for 3D shape synthesis (even accommodating some inconsistencies in the multi-view image).
\system makes 3D creation less like writing a perfect prompt or drawing a perfect shape and more like collaborating through
both mediums.

\bibliographystyle{ACM-Reference-Format}
\bibliography{references}

\end{document}